\documentclass[twocolumn,showpacs,showkeys,aps]{revtex4}
\usepackage{amsmath}

\begin{document}

\title{Towards a variational principle for motivated vehicle motion}

\author{Ihor Lubashevsky}
 \email{ialub@fpl.gpi.ru}
\affiliation{Theory Department, General Physics
Institute, Russian Academy of Sciences, Vavilov str., 38, Moscow, 119991,
Russia}
\author{Sergey Kalenkov}
\affiliation{Physics Department, Moscow State University for Technology
``MAMI'', Bol'shaya Semenovskaya str., 38, Moscow, 105831, Russia}
\author{Reinhard Mahnke}
\affiliation{Fachbereich Physik, Universit\"{a}t Rostock, D--18051 Rostock,
Germany}

\date{\today}

\begin{abstract}
We deal with the problem of deriving the microscopic equations governing the
individual car motion based on the assumptions about the strategy of driver
behavior. We suppose the driver behavior to be a result of a certain compromise
between the will to move at a speed that is comfortable for him under the
surrounding external conditions, comprising the physical state of the road, the
weather conditions, \textit{etc.}, and the necessity to keep a safe headway
distance between the cars in front of him. Such a strategy implies that a
driver can compare the possible ways of his further motion and so choose the
best one. To describe the driver preferences we introduce the priority
functional whose extremals specify the driver choice. For simplicity we
consider a single-lane road. In this case solving the corresponding equations
for the extremals we find the relationship between the current acceleration,
velocity and position of the car. As a special case we get a certain
generalization of the optimal velocity model similar to the ``intelligent
driver model'' proposed by Treiber and Helbing.
\end{abstract}

\maketitle

\section{Introduction}

The fundamentals of the traffic flow dynamics are far from being well
established because up to now it is not evident what the specific form of the
microscopic equations governing the individual car motion should be. The matter
is that the car motion is controlled by the motivated driver behavior rather
than obeys Newton's laws. In fact, the behavior of a driver is due to a certain
compromise between the will to move at a speed that is comfortable for him and
attained on the empty road, on one side, and the need to avoid possible
traffic accidents, on the other side. So, comparing the car ensembles with
physical systems it is not obvious beforehand that there is a direct
relationship between the acceleration of a given car and the positions and the
velocities of the other cars as is the case for physical particles. In
addition, the linear superposition typical in the interaction of physical
particles is not self-evident to hold for the vehicle interaction.

By contrast, on macroscopic scales the car ensembles exhibit a wide class of
critical and self-organization phenomena widely met in physical systems (for a
review see Ref.~\onlinecite{KL2,KL4,Sh00,Hrev}). It should be pointed out that
fish and bird swarms, colonies of bacteria, pedestrians, \textit{etc.} also
demonstrate similar cooperative motion (for a review see Ref.~\onlinecite{Hrev}
and also Ref.~\onlinecite{Ac1,Ac2,Ac3}). So the cooperative behavior of many
particle systems, including social and biological ones, seems to be of more
general nature then the mechanical laws and to find out what microscopic
regularities are responsible for the cooperative phenomena in the general case
is a challenge problem. The traffic dynamics is widely studied in this context
also due to the large potential for industrial applications.

The currently adopted approach to specifying the microscopic governing
equations of the individual car motion is the so-called social force model or
the generalized force model. Its detailed motivation and description can be
found in Ref.~\onlinecite{SFM1,SFM2,SFM3}, here we only touch its basic ideas.
At each moment $t$ of time a given driver $\alpha $ gets up or lets down the
speed $v_{\alpha }$ of his car, or keeps its value unchanged depending on the
road conditions and the arrangement in the neighboring cars:
\begin{equation}
\frac{dv_{\alpha }}{dt}=f_{\alpha }(v_{\alpha }) + \sideset{}{'}\sum_{\alpha '}
f_{\alpha \alpha^{\prime }}(x_{\alpha },v_{\alpha }\,|\,x_{\alpha ^{\prime
}},v_{\alpha ^{\prime }}). \label{int.1}
\end{equation}
Here the term $f_{\alpha }(v_{\alpha })$ taken typically in the form
\begin{equation*}
f_{\alpha }(v_{\alpha })=\frac{v_{\alpha }^{0}-v_{\alpha }}{\tau _{\alpha }}
\end{equation*}
describes the driver tendency to move on the empty road at a certain fixed
speed $v_{\alpha }^{0}$ depending on the physical state of the road, weather
conditions, the legal traffic regulations, \textit{etc}. The relaxation time
$\tau_{\alpha }$ characterizes the acceleration capability of the given car as
well as the delay in the driver control over the headway. The term $f_{\alpha
\alpha ^{\prime }}(x_{\alpha },v_{\alpha }\,|\,x_{\alpha ^{\prime }},v_{\alpha
^{\prime }})$ describes the interaction of car $\alpha $ with car $\alpha
^{\prime }$ ($\alpha ^{\prime }\neq \alpha $) that is due to the necessity for
driver $\alpha $ to keep a certain safe headway distance between the cars. The
force $f_{\alpha \alpha ^{\prime }}(x_{\alpha },v_{\alpha }\,|\,x_{\alpha
^{\prime }},v_{\alpha ^{\prime }})$ is assumed to depend directly only on the
velocities $v_{\alpha }$, $v_{\alpha ^{\prime}}$ and the position $x_{\alpha}$,
$x_{\alpha ^{\prime}}$ of this car pair and being of nonphysical nature does
not meet the third Newton's law, i.e. in the general case $f_{\alpha \alpha
^{\prime }}\neq - f_{\alpha ^{\prime }\alpha }$. Paper~\cite{GF} presents and
discusses possible generalized ansatzes of the dependence $f_{\alpha \alpha
^{\prime }}(x_{\alpha },v_{\alpha }\,|\,x_{\alpha ^{\prime }},v_{\alpha
^{\prime }})$.

Special cases of this model bear own names. In particular, for a single-lane
road when all the cars can be ordered according to the position on the road in
the car motion direction,
\begin{equation*}
\ldots <x_{\alpha -2}<x_{\alpha -1}<x_{\alpha }<x_{\alpha +1}<x_{\alpha
+2}<\ldots \,,
\end{equation*}
solely the interaction of the nearest neighboring cars $\alpha $ and $\alpha+1$
is taken into account, i.e. $f_{\alpha \alpha ^{\prime }}\neq 0$ for $\alpha
^{\prime }=\alpha +1$ and, may be, $\alpha ^{\prime }=\alpha -1$ only. For
this case Bando \textit{et al. }\cite{Bando1,Bando2} proposed the optimal
velocity model that describes the individual car motion as
\begin{equation}
\frac{dv_{\alpha }}{dt}=\frac{1}{\tau }\left[ \vartheta _{\text{opt}%
}(x_{\alpha +1}-x_{\alpha })-v_{\alpha }\right] \,,  \label{int.2}
\end{equation}
where $\vartheta_{\text{opt}}(\Delta )$ is the steady-state velocity (the
optimal velocity) chosen by drivers for the given headway distance $\Delta
=x_{\alpha +1}-x_{\alpha }$ between the cars. This model and its modification
successfully were used to explain the properties of the ``stop-and-go'' waves
developing in dense traffic on single-lane roads (see, e.g. Ref.~\onlinecite
{CF1,CF2,CF2a,CF3,CF4,CF5,CF6,CF7,CF8,CF9,CF10,CF11,CF12,CF13,CF14,CF15,CF16}).

However, on multilane highways the behavior of traffic flow becomes
sufficiently complex because of the strong correlations in the car motion at
different lanes (for a review see~Ref.~\onlinecite{KL2,KL4,Sh00,Hrev}). In this
case it is not sufficient to confine the consideration to the interaction
between the nearest neighboring cars only and one has to specify several
independent components of the social forces $\{f_{\alpha \alpha ^{\prime
}}(x_{\alpha },v_{\alpha }\,|\,x_{\alpha ^{\prime }},v_{\alpha ^{\prime }})\}$.
As a result the number of essential fitting parameters entering the social
forces increases substantially. A possible way to overcome this problem is to
formulate a mathematical principle characterizing the strategy of driver
behavior in terms of a certain functional quantifying the objectives pursued by
drivers. Constructing this functional may be also at the phenomenological level
can be lightened by the clear physical meaning of the driver objectives. Then,
using standard techniques one will derive the required governing equations. The
given work presents our first steps towards such an approach. It should be
noted that the derivation of microscopic governing equations for systems with
motivated behavior based on a certain ``optimal self-organization'' principle
was discussed in Ref.~\onlinecite{Hrev,Ac3,MSO1,MSO2,MSO3}. The main idea of
this approach is the assumption that individuals try to minimize the
interaction strength or, what is actually the same, to optimize their own
success and to minimize the efforts required for this.

Our approach is related to the concepts of mathematical economics, namely, to
the concepts of preferences and utility (see, e.g. Ref.~\onlinecite{Ut}). We
suppose that at each moment $t_{0}$ of time a driver plans his motion in a
certain way in order, first, to move as fast as possible and, second, to
prevent traffic accidents. In particular, in the previous paper~\cite{we} using
this idea we developed a model explaining in a simple way the experimentally
observed sequence of the first order phase transitions from the ``free flow''
to the jam phase through the ``synchronized mode'' \cite{K1,K2,K3}. Such a
strategy actually implies that a driver evaluates any possible path of his
further motion, $\{\chi (t),~t>t_{0}\}$, with respect to its preferability. In
other words, a driver can compare any two paths $\{\chi _{1}(t),~t>t_{0}\}$ and
$\{\chi _{2}(t),~t>t_{0}\}$ and decide which of them is more preferable, for
example, $\chi _{1}(t)$. The latter relation will be designated as $\chi
_{1}(t)\succ \chi _{2}(t)$. Obviously, the given relation exhibits the
transitivity, i.e. if $\chi _{1}(t)\succ \chi _{2}(t)$ and $\chi _{2}(t)\succ
\chi _{3}(t)$ then $\chi _{1}(t)\succ \chi _{3}(t)$. In this case we may seek
for a priority functional $\mathcal{L}\left\{\chi\right\}$ meeting the
condition $\mathcal{L}\left\{\chi_{1}\right\}
> \mathcal{L}\left\{\chi_{2}\right\}$ when and only when $\chi _{1}\succ \chi
_{2}$. Finally the driver chooses the best path $\chi_{\text{opt}}(t)$ of his
further motion maximizing the priority functional $\mathcal{L}\left\{ \chi
\right\}$. So its extremals have to satisfy the desired microscopic governing
equation of individual car motion. It should be noted that chosen path $\chi
_{\text{opt}}(t)$ of the planned motion specifies the acceleration at the
current time moment $t_{0}$ rather than the real trajectory $x(t)$ of the car
motion because at the next time $t>t_{0}$ the driver again plans his motion in
the same way, introducing the corrections caused by changes in the
surroundings. The same concerns the car velocity and its acceleration,
therefore to avoid possible misunderstanding we will designate the real
velocity and acceleration of the car as $v(t)$ and $a(t)$ whereas these values
corresponding to the optimal $\chi_{\text{opt}}(t)$ will be labelled by
$\{\nu(t),~t>t_{0}\}$ and $\{\varpi(t),~t>t_{0}\}$, respectively.

In this way the problem of specifying many independent components of the
social forces is replaced by constructing the priority functional describing
the driver compromise between the will to move as fast as being allowed by the
physical state of the road and the necessity to avoid possible traffic
accidents. So to obtain the priority functional we may apply to the general
assumptions about the driver behavior.

\section{Variational principle for the individual car motion}

At the first step we should determine the collection of phase variables
characterizing the quality of a given car motion. We note that for the driver
under consideration the neighboring car arrangement and its evolution should be
regarded as given beforehand. Indeed, it cannot be directly controlled by him
and, so, has to be treated as the external conditions. Formulating this
problem we actually assume the existence of a certain collection of variables
\textit{taken at the current time moment} $t$ that completely quantify the
priority measure of the car motion at the same time. Adopting the latter
assumption we may construct the priority functional $\mathcal{L}\left\{ \chi
\right\}$ in terms of a certain integral of a function $\mathcal{F}$ of the
phase variables with respect to time.

Applying to the conventional driver experience we will characterize the
individual car motion at each time moment $t$ by its position on the road
$x(t)$, the velocity $v(t)$, and the acceleration $a(t)$. For a multilane
highway, for example, the position of a car should also bear the information
about the lane occupied by the car, but this problem will be considered
elsewhere and in the present paper we confine our consideration to a
single-lane road only. In this property a car is distinct from a physical
particle because the motion of the latter is completely determined by the
current position and velocity. The variables $x(t)$, $v(t)$, and $a(t)$,
however, exhibit different behavior. The coordinate $x(t)$ and the velocity
$v(t)$ of the car vary continuously, i.e. the driver cannot change them
immediately. In contrast, the acceleration $a(t)$ may exhibit sharp jumps
because it is the acceleration that is controlled directly by the driver
without remarkable delay. In such an analysis it is quite reasonable to ignore
the short physiological delay in the driver behavior to changes in the
surroundings, allowing sharp jumps in the dependence $a(t)$. Therefore,
planning his further path of motion the driver regards the position
$x_0:=x(t_{0})$ and the velocity $v_0:=v(t_{0})$ of the car at the current
moment of time $t_{0}$ as the initial data.

Now let us write the general form of the priority functional
$\mathcal{L}\left\{\chi \right\}$ for a trial path $\{\chi(t),~t>t_{0}\}$ of
the further motion:
\begin{equation}
 \label{v.1}
\mathcal{L}\left\{ \chi \right\} =-\int\limits_{t_{0}}^{\infty }dt\,\exp \Bigl(
-\frac{\chi -x_0}{\ell }\Bigr) \mathcal{F}(t,\chi ,\nu ,\varpi ),
\end{equation}
where $\mathcal{F}(t,\chi ,\nu ,\varpi )$ is the density of the path priority
measure, $\nu :=d\chi /dt$ and $\varpi :=d^{2}\chi /dt^{2}$, and the
exponential cofactor reflects the fact that drivers can monitor the traffic
flow state and so plan the motion only inside a certain region of length $\ell$
in front of them. In addition, the leading minus has been chosen in order to
reduce finding the maximum of the functional $\mathcal{L}\left\{ \chi \right\}
$ to getting the minimum of integral~(\ref{v.1}) as it is the typical case in
physical theories. The direct dependence of the function
$\mathcal{F}(t,\chi,\nu,\varpi)$ on the time $t$ reflects the effect of the
surroundings, i.e. the physical road state and the neighboring car arrangement
on the driver planning.

According to the adopted assumption the driver chooses the path
$\chi_{\text{opt}}(t)$ of his further motion that maximizes the functional
$\mathcal{L} (t,\chi ,\nu ,\varpi )$ and at the current time $t_{0}$ together
with other all trial paths $\{\chi(t)\}$ meets the conditions
\begin{equation}
\chi (t_{0})=x_0,\quad \nu (t_{0})=v_0\,. \label{v.2}
\end{equation}
Besides, the present paper analyzes the car motion in the traffic flow, i.e.
does not consider any way by which a fixed car leaves the traffic flow, for
example, to stop. So we assume all the trial paths to exhibit bounded
variations, i.e. there are such constants $C_{\chi }^{l}\,$, $C_{\chi }^{u}\,$,
$C_{\nu }\,$, and $C_{\varpi }$ that
\begin{gather}
C_{\chi }^{l}(t-t_{0})<\chi (t)<C_{\chi }^{u}(t-t_{0}),\nonumber\\
\nu(t)<C_{\nu }\,,^{\quad }\left|\varpi (t)\right|<C_{\varpi }\,.\label{v.3}
\end{gather}
Then using the standard technique we get the governing equation for the
extremals of the priority functional $\mathcal{L}\left\{ \chi \right\}$
following from the condition $\delta \mathcal{L}\left\{ \chi \right\} =0$ at
$\chi (t)=\chi _{\text{opt}}(t)$:
\begin{widetext}
\begin{equation}
\frac{d^2}{dt^2}\left\{\exp\left[ -\frac{\chi}{\ell}\right] \frac
{\mathcal{F}(t,\chi,\nu ,\varpi )}{\partial \varpi }\right\}
-
\frac{d}{dt}\left\{ \exp \left[ -\frac{\chi}{\ell }\right] \frac{
\partial \mathcal{F}(t,\chi ,\nu ,\varpi )}{\partial \nu }\right\}
+
\frac{\partial}{\partial\chi}\left\{ \exp \left[-\frac{\chi}{\ell} \right]
\mathcal{F}(t,\chi ,\nu ,\varpi )\right\} =0\,. \label{v.4}
\end{equation}
By virtue of (\ref{v.3}) the function $\mathcal{F}(t,\chi ,\nu ,\varpi )$
exhibits bounded variations, which enables us to integrate equation~(\ref{v.4})
twice with respect to time $t$ reducing it to the following
\begin{multline}
\frac{\partial \mathcal{F}(t,\chi(t),\nu(t),\varpi(t))}{\partial \varpi(t)}  =
-\int\limits_{t}^{\infty }dt^{\prime }\exp \left[ -\frac{\chi (t^{\prime
})-\chi (t)}{\ell }\right] \frac{\partial \mathcal{F}(t^{\prime },\chi
(t^{\prime }),\nu (t^{\prime }),\varpi (t^{\prime }))}{\partial
\nu (t^{\prime })}\\
{}-\int\limits_{t}^{\infty }dt^{\prime }\int\limits_{t^{\prime }}^{\infty
}dt^{\prime \prime }\frac{\partial }{\partial \chi (t^{\prime \prime
})}\left\{ \exp \left[ -\frac{\chi (t^{\prime \prime })-\chi (t)}{\ell
}\right] \mathcal{F}(t^{\prime \prime },\chi (t^{\prime \prime }),\nu
(t^{\prime \prime }),\varpi (t^{\prime \prime }))\right\}\,. \label{v.5}
\end{multline}
\end{widetext}
Equation (\ref{v.5}) relates the planned acceleration $\varpi $ to the car
position $\chi $ and the velocity $\nu $. Subjecting this equation to the
initial conditions~(\ref{v.2}) we can find the optimal path $\chi _{
\text{opt}}(t)$ and then setting $t=t_{0}$ we will get the desired relationship
of the real current position $x(t_{0})$ and velocity $v(t_{0})$ of the car with
the acceleration $a(t_{0})$ that driver takes under these conditions. Exactly
the expression to be obtained in such a way is the microscopic governing
equation of the individual car motion. In addition, it should be noted that
equation~(\ref{v.4}) is of fourth order as it must because a trial path is
fixed in part by the position and the velocity at the initial and terminal
points. However, in the case under consideration the characteristics of the
terminal point are replaced by conditions~(\ref{v.3}), allowing us to reduce
the order.

Now let us demonstrate the proposed approach analyzing a simple example.

\section{The generalized optimal velocity model}

In constructing the priority functional we assumed that in the steady-state
traffic flow a driver prefers to move at a certain speed $\vartheta
_{\text{opt}}(t,x)$ depending on the surroundings and given beforehand.
Besides, we consider the car motion without acceleration to be the best way of
driving. Therefore we adopt the following ansatz:
\begin{equation}
\mathcal{F}(t,x,v,a)=\frac{1}{2}\left[ v-\vartheta _{\text{opt}}(t,x)\right]
^{2}+\frac{1}{2}\tau ^{2}a^{2},  \label{m.1}
\end{equation}
where the time scale $\tau $ characterizes the acceleration capability of the
given car. The driver monitoring the car arrangement in front of him can
predict the situation development, which is described in terms of the linear
dependence of the optimal velocity $\vartheta _{\text{opt}}(t,x)$ on the time
$t$ and the distance $x$:
\begin{equation}
\vartheta _{\text{opt}}(t,x)=\vartheta _{\text{opt}}^{0}\left[ 1+\varepsilon
_{t}\frac{t-t_{0}}{\tau }+\varepsilon _{x}\frac{x-x_{0}}{\ell }\right] ,
\label{m.2}
\end{equation}
where $\vartheta _{\text{opt}}^{0}=\vartheta _{\text{opt}}(t_{0},x_{0})$ and
$\varepsilon _{t}$, $\varepsilon _{x}$ are constants regarded here as small
parameters of the same order. In addition, the difference $v(t)/\vartheta
_{\text{opt}}^{0}-1$ is also assumed to be of order of $\varepsilon _{t}\sim
\varepsilon _{x}$. We have adopted the linear dependence of $\vartheta
_{\text{opt}}(t,x)$ on $t$ and $x$ because it seems quite reasonable that a
driver uses the linear approximation in estimating the position of the cars in
front of him.

Substituting (\ref{m.1}) and (\ref{m.2}) into (\ref{v.5}) and truncating all
the terms whose order exceeds $\varepsilon _{t}\sim \varepsilon _{x}\sim
(v(t)/\vartheta _{\text{opt}}^{0}-1)$ we get
\begin{widetext}
\begin{equation}
\tau ^{2}\varpi(t) +\int\limits_{t}^{\infty }dt^{\prime }\exp \left[ -\frac{
\vartheta _{\text{opt}}^{0}(t^{\prime }-t)}{\ell }\right] \left[ \nu
(t^{\prime })-\vartheta _{\text{opt}}^{0}\right] =\frac{\ell }{\tau \vartheta
_{\text{opt}}^{0}}\left( \varepsilon _{t}\ell +\varepsilon _{x}\tau \vartheta
_{\text{opt}}^{0}\right) \left[ 1+\frac{\vartheta
_{\text{opt}}^{0}(t-t_{0})}{\ell }\right] \,. \label{m.3}
\end{equation}
\end{widetext}
Multiplying equation (\ref{m.3}) by the factor $\exp (-\vartheta
_{\text{opt}}^{0}t/\ell )$, differentiating the obtained result with respect
to $t$, and solving the equation subject to the initial condition~(\ref{v.2})
we get the dependence $\nu (t)$ obeying (\ref{m.3}). The differentiating $\nu
(t)$ at the current moment $t_{0}$ of time point we obtain the desired
microscopic governing equation for the individual car motion:
\begin{equation}
\frac{dv}{dt}=-\frac{1}{\tau }\kappa \left( v-\vartheta
_{\text{opt}}^{0}\left[ 1+\kappa \left( \varepsilon _{t}+2\sigma \varepsilon
_{x}\right) \right] \right)\,, \label{m.4}
\end{equation}
where the constants
\begin{equation*}
\kappa =\left( \sigma +\sqrt{\sigma ^{2}+1}\right) ^{-1}\quad \text{and} \quad
\sigma =\frac{\tau \vartheta _{\text{opt}}^{0}}{2\ell }\,.
\end{equation*}
In particular, let the optimal velocity $\vartheta _{\text{opt}}(t,x)=\vartheta
_{\text{opt}}(\Delta )$ be specified completely by the headway distance $\Delta
=x_{\alpha +1}-x_{\alpha }$ between the given car $\alpha $ and the nearest one
$\alpha +1$ in front of it. Then within the linear approximation of the
situation development the driver of the car $\alpha$ can anticipate that the
headway distance will change in time as
$$
\Delta(t) = \Delta(t_0) + [v_{\alpha +1}(t_0)-v_{\alpha }(t_0)](t-t_0)\,.
$$
This expression together with the dependence $\vartheta _{\text{opt}}(\Delta )$
enables us to calculate the specific value of the constant $\varepsilon _{t}$
and then to rewrite formula~(\ref{m.4}) as
\begin{equation}
\frac{dv}{dt}=-\frac{1}{\tau }\kappa \left[ v-\vartheta _{\text{opt}}(\Delta
)-\kappa \tau \delta v\frac{d\vartheta _{\text{opt}}(\Delta )}{d\Delta
}\right]\,, \label{m.5}
\end{equation}
where we have introduced the relative velocity $\delta v=v_{\alpha
+1}-v_{\alpha }$ of the car $\alpha +1$ with respect to the given car $\alpha $
and omitted the argument $t_0$ assuming all the values to be taken at the
current moment of time.

It should be noted that the obtained expression (\ref{m.5}) is similar to the
phenomenological dependence $\vartheta _{\text{opt}}(\Delta ,\delta v)$
generalizing the standard optimal velocity model~(\ref{int.2}), the
``intelligent driver model'' proposed by Treiber and Helbing~\cite{IDM1} (see
also Ref.~\onlinecite{IDM2}).

\section{Conclusion}

Concluding the present paper we once more remember its main key points.

We deal with the problem of deriving microscopic equations governing the motion
of individual vehicles. The currently adopted approaches similar to the social
force model relate in the spirit of Newton's laws the acceleration of a given
car to the position and velocities of the neighboring cars. In order to apply
such models to analysis of the traffic dynamics one has to specify all the
essential components of the corresponding effective forces acting between the
cars. However when the vehicle interaction becomes sufficiently complex as it
is the case, for example, for dense traffic on multilane highways such an
approach meets the problem of large number of fitting parameters.

The present paper proposed a possible way to avoid the aforementioned
difficulty. The main idea is to describe at the first step the strategy of the
driver behavior determined by the compromise between the driver will to move as
fast as possible on the given road, on one hand, and the necessity to keep a
safe headway distance and not to interfere with cars moving at neighboring
lanes, on the other hand. This assumption actually implies that a driver can
compare various ways of his further motion with respect their relative
preference and choose the best (optimal) one at each time moment. This choice
gives the relationship between the acceleration of the car under consideration
and the arrangement of neighboring cars.

Following the concepts of mathematical economics we have introduced a priority
functional in order to quantify the driver choice. The extremals of this
priority functional describe the optimal path of the driver further motion. At
the present paper we have considered traffic flow on a single lane road,
constructed in this case the general form of the priority functional, and
derived the equations for its extremals corresponding to the car motion with
traffic flow. The latter means that here we do not analyze the paths by which a
car enters or leave traffic flow on the given road because the question
deserves an individual investigations.

By way of example, we have considered a special case leading to an expression
relating the current acceleration of a fixed car to the headway distance
between this car and one in front of it as well as their relative velocity. The
obtained equation turned out to be similar to the ``intelligent driver model''
by Treiber and Helbing.


\begin{thebibliography}{99}
\bibitem{KL2}
B.\,S. Kerner, in: \textit{Transportation and Traffic Theory}, edited by A.
Ceder (Pergamon, Amsterdam, 1999), p.~147.

\bibitem{KL4}
B.\,S. Kerner, in: \textit{Traffic and Granular Flow'99}, edited by D.~Helbing,
H.\,J.~Herrmann, M.~Schreckenberg and D.\,E.~Wolf (Springer-Verlag, Singapore,
2000), p.~253.

\bibitem{Sh00} D. Chowdhury, L. Santen, and A. Schadschneider, ``Statistical
physics of vehicular traffic and some related systems,'' \textit{Physics
Reports}, \textbf{329}, 199--329 (2000).

\bibitem{Hrev} D. Helbing, ``Traffic and related self-driven many-particle
systems,'' e-print: cond-mat/0012229.

\bibitem{Ac1} A. Czir\'{o}k and T. Vicsek, in: \textit{Traffic and Granular
Flow'97}, edited by M.~Schreckenberg and D.\,E.~Wolf (Springer-Verlag,
Singapore, 1998), p.~3.

\bibitem{Ac2} H. Helbing, in: \textit{Traffic and Granular Flow'97}, edited
by M.~Schreckenberg and D.\,E.~Wolf (Springer-Verlag, Singapore, 1998), p.~21.

\bibitem{Ac3}  H. Helbing, A. Czir\'{o}k, and T. Vicsek, in: \textit{Traffic
and Granular Flow'99}, edited by D.~Helbing, H.\,J.~Herrmann, M.~Schreckenberg
and D.\thinspace E.~Wolf (Springer-Verlag, Singapore, 2000), p.~147.

\bibitem{SFM1}  K. Lewin, \textit{Field Theory in Social Sciences} (Harper
and Brothers, New York, 1951).

\bibitem{SFM2}  D. Helbing, \textit{Quatitative Sociadynamics. Stochastic
Methods and Models of Social Interaction Processes} (Kluwer Academic,
Boston, 1995).

\bibitem{SFM3}  D. Helbing, \textit{Verkehrsdynamik} (Springer-Verlag,
Berlin, 1997).

\bibitem{GF}  D. Helbing and B. Tilch, Phys. Rev. E \textbf{58}, 133 (1998).

\bibitem{Bando1}  M. Bando, K. Hasebe, A. Nakayama, A. Shibata, and Y.
Sugiyama, Phys. Rev. E \textbf{51}, 1035 (1995).

\bibitem{Bando2}  M. Bando, K. Hasebe, K. Nakanishy, A. Nakayama, A.
Shibata, and Y. Sugiyama, J. Phys. Rev. I \textbf{5}, 1389 (1995).

\bibitem{CF1} T.\,S. Komatsu and S. Sasa, Phys. Rev. E \textbf{52}, 5574 (1995).

\bibitem{CF2}  K. Nakanishi, K. Itoh, Y. Igarashi, and M. Bando, Phys. Rev.
E \textbf{55}, 6519 (1997).

\bibitem{CF2a}  Y. Sugiyama and H. Yamada, Phys. Rev. E \textbf{55}, 7749
(1997).

\bibitem{CF3}  T. Nagatani, Physica A \textbf{248}, 353 (1998).

\bibitem{CF4}  H. Hayakawa and K. Nakanishi, Phys. Rev. E \textbf{57}, 3839
(1998).

\bibitem{CF5}  T. Nagatani and K. Nakanishi, Phys. Rev. E \textbf{57}, 6415
(1998).

\bibitem{CF6}  T. Nagatani, Phys. Rev. E \textbf{58}, 4271 (1998).

\bibitem{CF7}  M. Bando, K. Hasebe, K. Nakanishi, and A.Nakayama, Phys. Rev.
E \textbf{58}, 5429 (1998).

\bibitem{CF8}  M. Muramatsu and T. Nagatani, Phys. Rev. E \textbf{60}, 180
(1999).

\bibitem{CF9}  T. Nagatani, Phys. Rev. E \textbf{60}, 6395 (1999).

\bibitem{CF10}  P. Berg, A. Mason, and A. Woods, Phys. Rev. E \textbf{61},
1056 (2000).

\bibitem{CF11}  T. Nagatani, Phys. Rev. E \textbf{61}, 3534 (2000).

\bibitem{CF12}  T. Nagatani, Phys. Rev. E \textbf{61}, 3564 (2000).

\bibitem{CF13}  K. Nakanishi, Phys. Rev. E \textbf{62}, 3349 (2000).

\bibitem{CF14}  E. Tomer, L. Safonov, and S. Havlin, Phys. Rev. Lett.
\textbf{84}, 382 (2000).

\bibitem{CF15}  N. Mitarai and H. Nakanishi, Phys. Rev. Lett. \textbf{85},
1766 (2000).

\bibitem{CF16}  P. Berg and A. Woods, Phys. Rev. E \textbf{63}, 6107 (2001).

\bibitem{MSO1}  D. Helbing and P. Moln\'{a}r, in \textit{Self-Organization
of Complex Structures. From Individual to Collective Dynamics}, edited by
F.~Schweitzer (Gordon and Breach, London, 1997), p.~567.

\bibitem{MSO2}  D. Helbing and T. Vicsek, \textit{Self-Organized Optimality
in Driven Systems with Symmetrical Interaction}, e-print:~cond-mat/9903319.

\bibitem{MSO3}  D. Helbing and T. Vicsek, \textit{Optimal Self-Organization},
e-print:~cond-mat/9904327.

\bibitem{Ut}  \textit{Preferences, Utility, and Demand}, edited by
J.\,S. Chipman, L. Hurwicz, M.\,K. Richter, and H.\,F. Sonnenschein (Harcourt
Brace Jovanovich, New York, 1971).

\bibitem{we}  I.\,A. Lubashevsky and R. Mahnke, Phys. Rev. E 62,
6082 (2000).

\bibitem{K1}  B.\,S. Kerner, Phys. Rev. Lett. \textbf{81}, 3797
(1998).

\bibitem{K2}  B.\,S. Kerner and H. Rehborn, Phys. Rev. E \textbf{53},
R1297 (1996).

\bibitem{K3}  B.\,S. Kerner and H. Rehborn, Phys. Rev. Lett.
\textbf{79}, 4030 (1997).

\bibitem{IDM1}  M. Treiber and D. Helbing, \textit{Explanation of observed
features of self-organization in traffic flow}, e-print: cond-mat/9901239.

\bibitem{IDM2}  M. Treiber, A. Hennecke, and D. Helbing, in: \textit{Traffic
and Granular Flow'99}, edited by D.~Helbing, H.\,J.~Herrmann, M.~Schreckenberg
and D.\,E.~Wolf (Springer-Verlag, Singapore, 2000), p.~365.
\end{thebibliography}
\end{document}